\documentclass[prb,%
twocolumn,%
amsfonts,amsmath,amssymb,%
showpacs,%
floatfix,%
]{revtex4}

\usepackage{bm}
\usepackage{graphicx}
\usepackage{tabularx}
\usepackage{longtable}

\begin{document}
\title[%
Nonlinear wave-wave interactions in stratified flows: DNS
]{%
Nonlinear wave-wave interactions in stratified flows:
Direct numerical simulations
}
\author{%
Yuri V. \surname{Lvov}
}
\affiliation{
Department of Mathematical Sciences,
Rensselaer Polytechnic Institute,
Troy, New York 12180 USA
}

\author{Naoto \surname{Yokoyama}}
\affiliation{
Department of Mechanical Engineering,
Doshisha University,
Kyotanabe, Kyoto 610-0394 Japan
}
\thanks{Author to whom should be addressed}
\email{nyokoyam@mail.doshisha.ac.jp}

\date{\today}

\begin{abstract}
To investigate the formation mechanism of energy spectra of internal waves in the oceans, direct numerical
simulations are performed. The simulations are based on the reduced dynamical equations of 
rotating stratified turbulence. In the reduced dynamical equations
only wave  modes are retained, 
and vortices and horizontally uniform vertical shears are excluded.
Despite the simplifications, our simulations reproduce some key features of 
oceanic internal-wave spectra: accumulation of energy at near-inertial waves
and realistic frequency and horizontal wavenumber dependencies. 
Furthermore, we provide evidence that formation of the energy spectra in the inertial
subrange is dominated by scale-separated interactions with the
near-inertial waves.
These findings support observationally-based intuition that spectral energy density 
of internal waves is the result of predominantly wave-wave interactions.
\end{abstract}

\maketitle

\section{Introduction\label{sec:introduction}}
Oceanic internal waves are the waves whose restoring force is buoyancy in
stratified fluid.
These waves are excited
by flows over topography, tides and atmospheric disturbances.  The energy of the waves is then
transferred by nonlinear interactions among wavenumbers from large
scales to small scales, and is dissipated in the small spatial scales
by wave breaking.
Internal waves play a significant role in the general
circulation of oceans and hence the climate of the Earth.

Energy spectra of internal waves are vast with horizontal
wavelengths varying from $10$ meters to $10^5$ meters, vertical
wavelengths from $10$ meters to $10^3$ meters, and time periods from
$10^3$ seconds to $10^5$ seconds.

The complexity of the internal-wave fields arises not only from its extended
range of scales, but also from their interactions with the other major
players in ocean dynamics including eddies, mean currents and shear
flows. The main dynamical role of the internal waves is to store energy
and transfer it across different scales and large distances.  Hence
the waves constitute a large and complex geosystem containing a broad
range of interacting scales and affecting significantly most of the
active players in ocean dynamics.

However, surprisingly and despite all the complexities, energy spectra
of the internal waves in the oceans appear to be somewhat universal. It is 
given by the Garrett--Munk (GM) spectrum~\cite{GM72,GM75,GM_ARF}.
It is believed that the internal-wave spectrum
may be a result predominantly, if not exclusively, of nonlinear
interactions among waves.

In this paper we test the hypothesis by direct numerical simulations 
of the reduced model of stratified rotating turbulence. Our model contains 
only wave modes of stratified wave turbulence, and
completely excludes vortices and horizontally uniform vertical shears.
Despite these simplifications,
our numerical model reproduces some
key features of spectral energy density observed in the oceans. 

As a result of appearance of the GM spectrum, internal waves in the 
ocean have been a subject of active research ever since. An important 
milestone was a review by \citet{muller1986nia}. It focuses on
the resonant wave-wave interaction theory,  called the weak
turbulence theory. The result of the review is that the resonant
wave-wave interactions in the stratified wave turbulence are dominated
by specific, ``named''  nonlocal wave-wave interactions in the wavenumber space.
The classification of the ``named'' nonlocal interactions appeared first 
in \citet{mccomas-1977-7}.

Since then
it was understood that in the oceans local spectral energy density may change owing to
the propagation of energy from other parts of the ocean
as well as owing to the nonlinear wave-wave interactions.
A useful and intuitive
diagram in the wavenumber space that separates the two regions appear in
\cite{dasaro1991sia}.
More recently, \citet{levine32mgm} proposed modification of the GM
spectra to take into account the dependence of the characteristic
depth as a function of frequency $\omega$. Furthermore, \citet{lvov-2004-195}
developed a novel Hamiltonian structure for waves in stratified flows
that we use below for our numerical modeling. Historical observational data were
reviewed in \citet{lvov-2004-92}, where major
deviations from the GM spectrum were categorized. 
Useful phenomenological characterization of fluxes of energy in internal waves were put forward by \citet{polzin-2004}.  
 
However, the GM spectrum stood the test of time and still stands as the 
canonical model of the internal-wave spectral energy density. 
The GM spectrum 
is a function of
the frequency $\omega$ and the vertical
wavenumber $m$, $E(\omega,m)$. 
However, it is rather hard and expensive to measure the spectrum as a
function of both $\omega$ and $m$.
At least in the 1970's, when the series of the GM spectra are published \cite{GM72,GM75,GM_ARF},
only one-dimensional spectra were generally available (with the exception of the IWEX experiment). In particular, 
$\overline{E}_{\mathrm{time}}(\omega)$ was obtained from time series 
of mooring current meters and $\overline{E}_{\mathrm{vertical}}(m)$ from vertical profilers. 
Furthermore the large wavenumber limit of 
$\overline{E}_{\mathrm{vertical}}(m)$ has $m^{-2}$ dependency.

Garrett and Munk proposed that the spectrum is {\it separable}, that is the 
product of a function of $\omega$ and a function of $m$:
\begin{eqnarray}
E(\omega,m) \propto \overline{E}_{\mathrm{time}}(\omega) \overline{E}_{\mathrm{vertical}}(m).
\label{Separable}
\end{eqnarray}
Then $\overline{E}_{\mathrm{time}}(\omega)$ and $\overline{E}_{\mathrm{vertical}}(m)$
were properly normalized and chosen so that the resulting spectrum 
matches experimentally measured one-dimensional $\omega$ and $m$ spectra. 
The resulting spectrum is given by Eq.~(\ref{GM}) below.

The assumption of the separability allows one to construct a function of two arguments, 
$E(\omega,m)$ out 
of two one-dimensional functions, 
$\overline{E}_{\mathrm{time}}(\omega)$ and $\overline{E}_{\mathrm{vertical}}(m)$.
However, if one relaxes the assumption,
there is more
than one way to obtain the two-dimensional spectrum, $E(\omega, m)$, to fit the observed
one-dimensional spectra, $\overline{E}_{\mathrm{time}}(\omega)$ and
$\overline{E}_{\mathrm{vertical}}(m)$.  Moreover, it recently became
quite apparent that the assumption (\ref{Separable}) is not satisfied
in the oceans \cite{kurtPC}.
Our numerical simulations also demonstrate that the assumption
(\ref{Separable}) is not satisfied uniformly.
It also appears in our direct
numerical simulations that resulting energy spectra do not display
universal behavior. 
Rather, our simulations demonstrate accumulation of energy around the
horizontally longest waves.
Furthermore, we argue based on our direct numerical simulations that the
energy spectrum in the inertial subrange is determined by nonlocal interactions with the accumulation.
The nonlocal interaction in the wavenumber space is one of the ``named'' nonlocal interactions identified  by \citet{mccomas-1977-7}.
As a result of the nonlocal interactions, 
the details of behavior of the stratified wave turbulent
system depends upon details of the accumulation. Consequently
the resulting energy spectra are non-universal in our numerical experiments.

Despite apparent non-universality and non-separability 
of spectral energy density in the inertial subrange, our simulations do exhibit certain 
key features that are observed in the ocean. Namely, our simulations have clear 
peaks at inertial frequencies which correspond to the accumulation of energy,
as observed in the ocean.  
Our largest numerical simulation does demonstrate the $\omega^{-2}$ dependence of the energy spectrum that
appears prominently in moored observations. 
Furthermore our simulation demonstrate 
realistic $k^{-2}$ dependence on the horizontal wavenumbers. 
The behavior of the spectra in the inertial subrange, formation of accumulation, 
apparent non-universality and violation of separability (\ref{Separable}) 
can be qualitatively interpreted in terms of
the nonlocal ``named'' interactions.
These findings support observationally-based intuition that spectral energy density of 
internal waves is formed primarily by the nonlinear wave-wave interactions.

The stratified rotating {\em wave\/} turbulence is a
subset of a much more complicated subject of 
rotating stratified strong turbulence governed by the Navier--Stokes equation.
Complexities of the rotating stratified strong turbulence appear
owing to coexistence of waves, shears and vortices and their interactions.
The rotating and stratified strong turbulence has been a subject of intensive research in last few
decades. Extensive numerical studies of rotating turbulence, stratified turbulence and 
turbulence with both rotations and stratification were performed 
\cite{winters-1997,smith2002gsl,PhysRevE.68.036308,smith2005nra}. 
Accumulation of energy at 
the horizontally largest scales were reported in direct numerical
simulations~\cite{smith2002gsl,smith2005nra}. 
The accumulation in the rotating strong turbulence
happens presumably owing to the inverse cascade of two-dimensional turbulence.
In contrast, in the stratified rotating turbulence,
the resonant wave-wave interactions commonly cause the accumulation of energy
at the horizontal largest scales.

We emphasize that the rotating stratified wave turbulence is dominated by 
nonlocal wave-wave interactions in the wavenumber space. 
Anisotropic wave turbulence systems often exhibit nonlocal interactions:
drift wave turbulence, Rossby waves and MHD turbulence \cite{nazarenko1991niz,nazarenko2001nlm}.
The scenario is in contrast to the local interactions
of isotropic wave turbulence systems.
In the locally interacting systems the spectra are
insensitive to the details of large-scale and small-scale motions, and are the results
of the local interactions among wavenumbers. Examples of the
 locally interacting systems include waves on water surfaces.  In particular,
universal behavior was observed in direct numerical simulations of
gravity-wave and capillary-wave systems
\cite{kz_PRL_2002,pushkarev,naoto_jfm,lvov2006dai,dyachenko2004wtk}.
Note that isotropic Navier--Stokes turbulence is also widely believed to be a locally interacting system.

The paper is written as follows. In Sec.~\ref{HamiltonianSection}
we give the detailed description of our numerical model and
assumptions used. In Sec.~\ref{section:DNS} we elaborate on our
numerical methods, and explain pumping and damping mechanisms. In
Sec.~\ref{Results} we account for results of our numerical
experiments.
The formation mechanism of energy spectra is discussed in Sec.~\ref{sec:discussion}.
Section~\ref{sec:summary} provides summary.

\section{Hamiltonian formalism for internal waves}
\label{HamiltonianSection}
In this section we provide a description of the model that we use for our numerical 
study. The model is based on the Hamiltonian description of the wave modes 
of the incompressible stratified rotating flows in hydrostatic balance.
The Hamiltonian description appeared in~\citet{lvov-2004-195} and is presented here 
for completeness. 
Our model explicitly excludes vortices and horizontally uniform shear flows. 
Despite the simplification, the resulting spectrum does display some 
key features that are observed in the ocean. 

As a starting point, we take the equations of motion satisfied by an
incompressible stratified rotating flow in hydrostatic balance
under the Boussinesq approximation:
\begin{eqnarray}
\frac{\partial}{\partial t} \frac{\partial z}{\partial \rho} + \nabla \cdot \left(\frac{\partial z}{\partial \rho} \bm{u} \right) &=& 0 , \nonumber \\
\frac{\partial \bm{u}}{\partial t} +f \bm{u}^\perp+ (\bm{u} \cdot \nabla)
 \bm{u} + \frac{\nabla M}{\rho_0} &=& 0 ,
\nonumber
\\
\frac{\partial M}{\partial \rho} - g z &=& 0 .
\label{PrimitiveEquations}
\end{eqnarray}
These equations are derived from
mass and horizontal-momentum conservations and hydrostatic balance.
The equations are written in isopycnal coordinates with the density
$\rho$ replacing the height $z$ in its role as independent vertical
variable.  Here $\bm{u} = (u,v)$ is the horizontal component of the
velocity field,
and $\bm{u}^{\perp} = (-v,u)$.
The gradient operator
$\nabla = (\partial/\partial x, \partial/\partial y)$
acts along isopycnals.
The inertial frequency due to the rotation of the Earth $f$ is assumed to be constant,
$g$ is the acceleration of gravity,
and $\rho_0$ is a reference density (in its role as
inertia) which is taken to be a constant in the Boussinesq approximation.
Finally $M$ is the Montgomery potential
$$M=P+g\,\rho\,z \, .$$

The expression for the potential vorticity in these coordinates is,
\begin{equation}
 q = \frac{f+\nabla^{\perp} \cdot \bm{u}}{\Pi},
\label{PVorig}
\end{equation}
where $\nabla^{\perp} = (-\partial/\partial y, \partial/\partial x)$ is the two-dimensional rotation operator.
Here we introduced
$$\Pi = \frac{\rho}{g} \frac{\partial^2 M}{\partial \rho^2} = \rho \frac{\partial z}{\partial \rho} $$
to be a normalized differential layer thickness. The potential
vorticity is conserved along particle trajectories. Since the fluid
density is also conserved along the trajectories, an initial profile where the
potential vorticity is a function of the density will be preserved by
the flow.  Hence any initial chosen profile will stay in the fluid
all the time.  This observation allows us to propose that
\begin{equation}
  q = q_0(\rho) = \frac{f}{\Pi_0(\rho)} \, .
\label{PV}
\end{equation}
Here
$\Pi_0(\rho) $ is a reference stratification profile defined
as
\begin{eqnarray}
  \Pi_0(\rho) = - \frac{g}{N(\rho)^2} \, \label{PV2}
\end{eqnarray}
and
$N(\rho)$ is the buoyancy (Brunt--V\"{a}is\"{a}l\"{a}) frequency, which we shall regard
as a constant $N_0$.

In order to separate wave and vorticity dynamics, we decompose the
fluid velocity into its gradient and rotational parts, i.e.
\begin{eqnarray}
 \bm{u} = \nabla \phi + \nabla^{\perp} \psi \, .
\end{eqnarray}
The vorticity is derived from the potential vorticity
and is distinct from the vorticity in Cartesian coordinates.
In terms of the potentials $\phi$ and $\psi$, the constrain (\ref{PV}) reads
\begin{displaymath} f + \Delta \psi = q_0 \Pi \, .\nonumber \end{displaymath}
Therefore we can express $\psi$ as a function of $\Pi$ so that
Eqs.~(\ref{PV}) and (\ref{PV2}) are satisfied.    
As a result, if we redefine $\Pi$ as $\Pi+\Pi_0$, Eqs.~(\ref{PrimitiveEquations})
reduce to the pair:
\begin{eqnarray}
&&
 \frac{\partial \Pi}{\partial t} + \nabla \cdot \left((\Pi + \Pi_0) \left(\nabla \phi + \nabla^{{\perp}} \Delta^{-1}
 \left(q_0 \Pi-f\right)\right)\right)
  = 0 \, , \nonumber\\
&&
\frac{\partial \phi}{\partial t}  + \frac{1}{2}
  \left|\nabla\phi+ \nabla^{{\perp}}\Delta^{-1} \left(q_0 \Pi-f\right)
  \right|^2
  \nonumber \\
&&
\qquad
 +
   \Delta^{-1} \nabla \cdot
   \left( q_0 \Pi \,
    \left(\nabla^{{\perp}}\phi -
    \nabla\Delta^{-1}\left(q_0 \Pi-f\right)\right) \right)
  \nonumber \\
&&
\qquad
    + \frac{1}{\rho}
    \int^{\rho}\int^{\rho_2} \frac{\Pi(\rho_1)}{\rho_1}
    d\rho_1 d\rho_2 = 0 \,
. \label{LIWHAM}
\end{eqnarray}

The pair of equations~(\ref{LIWHAM})
form a canonical conjugate pair of the Hamiltonian equations,
\begin{equation}
 \frac{\partial \Pi}{\partial t}  = - \frac{\delta {\cal H}}{\delta \phi} \, ,
  \qquad
\frac{\partial \phi}{\partial t}  = \frac{\delta {\cal H}}{\delta \Pi} \, .
  \label{Canonical}
\end{equation}
The Hamiltonian is the sum of kinetic and potential energies,
\begin{eqnarray}
 {\cal H} &&
\!\!\!\!\!\!
 = \int
%
 d \bm{x} d \rho
 \left(
  \frac{g}{2} \left|\int^{\rho} d\rho^{\prime} \frac{\Pi(\bm{x}, \rho^{\prime})}{\rho^{\prime}}  \right|^2
\right.
\nonumber\\
&&
\!\!\!\!\!\!\!\!\!\!\!\!\!\!\!\!\!\!
\left.
 - \frac{1}{2} \left(
\Pi(\bm{x}, \rho) + \Pi_0 \right)
  \left|\nabla \phi(\bm{x}, \rho) + \nabla^{{\perp}}\Delta^{-1} q_0 \Pi(\bm{x}, \rho)
  \right|^2
  \right)
.
  \label{HTL2}
\end{eqnarray}

We therefore consider in this paper the following reduced model
with the following assumptions and constraints:
\begin{itemize}
\item only wave motions are considered
\item vortex motions are excluded
\item horizontally uniform vertical shear is excluded
\item potential vorticity is constrained to be constant on an isopycnal surface.
 Its value is  determined by the underlying rotations
\item constant underlying rotation $f=\mathrm{const}$ is assumed. 
Thus we explicitly exclude the $\beta$ effect.
\item hydrostatic balance is assumed.
\item buoyancy frequency $N(z)=\mathrm{const}$ is assumed to be constant with depth.
\end{itemize}

We do realize that
the oceans are more complicated than these idealizations.
Certainly for general ocean modeling the above assumptions constitutes
a gross simplification.
The reason for our choice is that we would like to find out whether 
it is sufficient to study wave-wave interactions alone to 
determine the form of the internal-wave spectral energy density. 
We show below that this is indeed the case to a large extent. 
We show that our reduced model reproduces key characteristic behavior of the oceans.
The effects of relaxing
the above assumptions is the subject of future work.

\section{Details of numerical methods: Implementation and interpretation}
\label{section:DNS}
\subsection{Wave turbulence}
To proceed, we
perform Fourier transformation and canonical transformation to
the field variable, $a(\bm{p})$, defined as
  \begin{eqnarray}
   a(\bm{p}) = \sqrt{\frac{\omega}{2 g}} \frac{N_0}{|\bm{k}|} {\widetilde
\Pi}(\bm{p})
    - i \sqrt{\frac{g}{2 \omega}}\frac{|\bm{k}|}{N_0} {\widetilde
\phi}(\bm{p}) \, ,
  \end{eqnarray}
with linear coupling of the Fourier components of the stratification profile,
${\widetilde \Pi}$,
and the horizontal velocity potential, ${\widetilde \phi}$.
The three-dimensional wavenumber, $\bm{p}$, consists of
a two-dimensional horizontal wavenumber in the isopycnal surface, $\bm{k}$,
and a vertical density wavenumber, $m$.
The linear frequency in the isopycnal coordinates is given by the dispersion relation,
  \begin{eqnarray}
   \omega(\bm{p}) = \sqrt{f^2 + \frac{g^2}{\rho_0^2 N_0^2}
\frac{|\bm{k}|^2}{m^2}} \, .
    \label{eq:dispersion}
  \end{eqnarray}
The usual vertical wavenumber, $k_z$, and the density wavenumber, $m$, are
related as
$m = - g/(\rho_0 N_0^2) k_z$.

Then, the pair of canonical equations of motion (\ref{Canonical}) is rewritten as a single 
canonical equation,
\begin{eqnarray}
    i \frac{\partial a(\bm{p})}{\partial t} = \frac{\delta {\cal
H}}{\delta a^{\ast}(\bm{p})}\label{eq:canonical}
\end{eqnarray}
 with the standard Hamiltonian of three-wave interacting systems \cite{zak_book},
\begin{eqnarray}
&& {\cal H} =  \int d\bm{p} \: \omega(\bm{p}) |a(\bm{p})|^2
   \nonumber\\
&&
 + \int \! d\bm{p} d\bm{p}_1 d\bm{p}_2
   \left(
   \left(
   V_{\bm{p}_1,\bm{p}_2}^{\bm{p}} a(\bm{p}) a^{\ast}(\bm{p}_1)
a^{\ast}(\bm{p}_2) + \mathrm{c.c.} \right)
\right.
   \nonumber\\
&&
\qquad\qquad\qquad
\left.
   +
   \left(
   U_{\bm{p},\bm{p}_1,\bm{p}_2} a(\bm{p}) a(\bm{p}_1) a(\bm{p}_2) +
\mathrm{c.c.}\right)
   \right)
\, .\nonumber \\
\label{eq:hamiltonian}
\end{eqnarray}
Here, $\delta/\delta a^{\ast}$ is the functional derivative
with respect to $a^{\ast}(\bm{p})$, which is the complex conjugate of $a(\bm{p})$,
and the abbreviation c.c.\ denotes complex conjugates.
The matrix elements, $V_{\bm{p}_1,\bm{p}_2}^{\bm{p}}$ and
$U_{\bm{p},\bm{p}_1,\bm{p}_2}$, have exchange symmetries such that
  $V_{\bm{p}_1,\bm{p}_2}^{\bm{p}} = V_{\bm{p}_2,\bm{p}_1}^{\bm{p}}$
and
  $U_{\bm{p},\bm{p}_1,\bm{p}_2} = U_{\bm{p},\bm{p}_2,\bm{p}_1} =
U_{\bm{p}_1,\bm{p},\bm{p}_2}$ \cite{lvov-2004-195}.

The Hamiltonian (\ref{eq:hamiltonian}) is the canonical form
of the Hamiltonian of wave turbulence system dominated by three-wave
interactions~\cite{zak_book}. The first term describes linear noninteracting waves, 
the second term correspond to the nonlinear three-wave scattering processes.  
The wave turbulence theory provides a powerful framework to describe spectral energy transfer 
in the 
systems dominated by wave-wave interactions. A detailed review of the wave turbulence theory and its applications to
internal waves is outside of the scope of the present paper, and is
given in \citet{iwthLPTN}. Here it is sufficient to note that there are the
following important classes of scale-separated resonant interactions among waves
\cite{mccomas-1977-7}:
\begin{itemize}
\item
The vertical backscattering of a high frequency wave by a low frequency wave of
twice the vertical wavenumber
into a second high frequency wave of oppositely signed vertical wavenumber.
This type of scattering is called Elastic Scattering (ES).
\item
The scattering of a high frequency wave by a low frequency and small vertical wavenumber wave
into a second, nearly identical high frequency and large vertical wavenumber wave. This type of scattering is called Induced Diffusion (ID).
\item
The decay of a small wavenumber wave into two large vertical wavenumber
waves of approximately one-half the frequency. This is called
Parametric Subharmonic Instability (PSI).
\end{itemize}

The classification provides a useful interpretive framework to
characterize resonant wave-wave interactions in stratified flows. 
We will show below that results of 
our numerical simulations can be qualitatively
interpreted by using this classification.
Detailed
theoretical analysis of scale-separated interactions of this type 
will be presented in \citet{iwthLPTN}.  

\begin{table*}
  \caption{Numerical parameters.
 The wavenumbers, $\bm{k}$ and $m$, are discretized
 and they have integer values.
 }
  \label{table:parameters}
 \begin{center}
\begin{ruledtabular}
\begin{longtable}{cccccc}
  & modes     & $f$ ($\times 10^{-4}$rad/sec) & $L_{\mathrm{v}}$ ($\times 10$kg/m$^3$)& forcing & initial condition\\
\hline
 I & $512^2 \times 256$  & $\sqrt{2}/3$  & 2.7 & none & GM\\
 II & $256^3$             & $0.25$        & 5   & $|\bm{k}|^2  + m^2  \leq 6^2$ & white noise\\
 III  & $256^3$             & $1$           & 5   & $|\bm{k}|^2  + m^2  \leq 6^2$ & white noise\\
 IV & $512^2 \times 256$  & $\sqrt{2}/3$  & 2.7 & none & GM without long waves \\
 V & $1024^2 \times 512$ & $\sqrt{2}/3$  & 2.7 & $\omega \sim 3f$ & white noise
\end{longtable}
\end{ruledtabular}
 \end{center}
\end{table*}

\subsection{Numerical setting}
To achieve non-equilibrium statistically (near-)steady states,
we have to model both processes of pumping
energy to the internal-wave field and of damping energy
from the field. The processes of the pumping include
interactions with surface waves and tides. How to model 
these processes in the wavenumber space is the subject of present
oceanographic research.  We model the pumping processes phenomenologically.
In what follows we assume that the pumping occurs on large length
scales, and is relatively local in the wavenumber space.
The processes that remove energy from the wave field include wave
breaking, turbulent dissipation, reflection from surface and bottom
boundary layers and interaction with topography. Again, the spectral
details of the processes is a subject of intensive research.  We
assume that the processes are especially effective for small length
scales (large wavenumbers).

In our numerical simulations,
most of the energy provided by the external forcing accumulates around the horizontally longest waves
owing to nonlinear interactions among waves.
The horizontally longest waves have frequencies near the inertial 
frequency $f$. Therefore the waves are called ``near-inertial waves.''
Then the energy is transferred through the inertial subrange, where there is no
significant pumping or damping.
Note that the inertial subrange therefore refers to the range in
wavenumber space without effective forcing and damping,
while the near-inertial waves refers to the waves with the near-inertial frequency.
Subsequently
energy is absorbed in the dissipation range.
It is the nonlinear interactions among internal waves that determine the form of the
spectrum in the inertial subrange and the formation of accumulation of energy at the near-inertial waves. The nonlinear interactions among waves is the main focus of the present paper.

In the direct numerical simulations, we add external forcing and
hyper-viscosity to the canonical equation (\ref{eq:canonical}).
Thus the dynamic equation used in the simulations is given by
\begin{eqnarray}
 \frac{\partial a(\bm{p})}{\partial t} = - i \, \omega(\bm{p}) a(\bm{p})
 + {\cal N}(a(\bm{p})) + F(\bm{p}) - D(\bm{p}) a(\bm{p}) \, . \label{EqnOfMotion}
\end{eqnarray}
Details of numerical algorithm are the following:
The linear terms,
i.e. a linear dispersion term and a dissipation term, $-D(\bm{p}) a(\bm{p})$,
are explicitly calculated.
The nonlinear terms, ${\cal N}(a(\bm{p}))$, are derived from the nonlinear parts of the canonical equation (\ref{eq:canonical}) with Hamiltonian (\ref{eq:hamiltonian}).
They are obtained numerically by a pseudo-spectral method with the phase shift
under the periodic boundary conditions for all three directions.
The external forcing, $F(\bm{p})$, is implemented
by fixing the amplitudes of several small wavenumbers to be constant in time.
This is commonly used to simulate forcing in numerical experiments.
The dissipation is modeled as hyper-viscosity:
\begin{equation}
D(\bm{p}) = D_{\mathrm{h}} |\bm{k}|^8 + D_{\mathrm{v}} |m|^4.
\label{hyper}
\end{equation}
Here, $ D_{\mathrm{h}}$ and $D_{\mathrm{v}}$ are chosen
so that the dissipation is effective for wavenumbers
larger than the half of maximum wavenumbers.

The wavenumbers are discretized as $\bm{p} = (2\pi/ L_{\mathrm{h}} \bm{k}, \: 2\pi/ L_{\mathrm{v}} m)$, where $L_{\mathrm{h}}$ and $L_{\mathrm{v}}$ are
horizontal periodic length and vertical period in the isopycnal coordinates,
and $\bm{k}$ and $m$ are integer-valued wavenumbers. We are going to 
use integer-valued (dimensionless) wavenumbers from here on in this paper. 
Time-stepping is implemented with the fourth-order Runge--Kutta method.
In all the simulations,
the buoyancy frequency and horizontal period are fixed at $N_0 = 10^{-2}$rad/sec and $L_{\mathrm{h}}=10^5$m, respectively.

We perform a series of five numerical experiments that are listed in Table~\ref{table:parameters}.
The total energy per unit periodic box of all the numerical experiments except Run V
is around $3 \times 10^3$J/(kg $\cdot$ m$^2$) which is characteristic of the oceans.
The total energy in Run V is around $1.2 \times 10^3$J/(kg $\cdot$ m$^2$).
The values of total energy density and the dissipation rate of Run V in Cartesian coordinate
are $1.3 \times 10^{-3}$J/kg and $5.0 \times 10^{-11}$W/kg, respectively.
These values  are in good agreement with observations and theories
\cite{winters-1997}.

Stratified rotating turbulence can be characterized by two dimensionless numbers,
the Rossby number $Ro$, which is the ratio of the inertia force to the Coriolis force,
and Richardson number $Ri$, which is the ratio of the buoyancy to the shear.
They are defined as
\begin{eqnarray}
Ro &=& \langle | (\bm{u} \cdot \nabla) \bm{u}| / |f \bm{u}^{\perp}| \rangle ,
\nonumber \\
Ri &=& \langle N^2 / S^2 \rangle ,
\nonumber
\end{eqnarray}
where $\langle \cdot \rangle$ denotes averaging in the numerical box
and $N$ and $S$ show local buoyancy frequency given by local stratification and local shear, respectively.
The values of the Rossby number and the Richardson number measured in Run V are
$2.7 \times 10^{-1}$ and $8.6 \times 10^2$, respectively.

\subsection{Integrated and cross-sectional spectral energy density}

The present paper is concentrated on the investigation of the behavior of 
spectral energy density in the inertial subrange.
As mentioned above our numerical scheme employs the pseudo-spectral
algorithm. Consequently it is convenient to concentrate our
attention on the $(k,m)$ spectra. Indeed, the numerical box is a
triple periodic box in the wavenumber space.  On the other hand,
oceanographers find it convenient to think in $(\omega,m)$ space.
Indeed, the $\omega$ and $m$ spectra could be measured
experimentally. 

There are merits to both ways
of thinking. For example, we will demonstrate below that the spectrum is
more separable in $(k,m)$ space for Run II and III. On the other
hand, Run V could be better interpreted in $(\omega,m)$ space.  
The oceanic spectra also tend to be more separable in $(\omega,m)$ space.  
Indeed, the GM spectrum is assumed to be separable in $(\omega,m)$ space.
We will use both ways of thinking through the paper.

Two-dimensional energy spectra are measured in the shell of radius $k$ as
\begin{eqnarray}
E(k, |m|) = \sum_{k- 1/2 \leq |\bm{k}^{\prime}| < k+ 1/2}
 \sum_{s=\pm 1} \omega |a(\bm{k}^{\prime}, s m)|^2 \, .
\end{eqnarray}
Integrated energy spectra are defined as
\begin{equation}
\overline{E}_{\mathrm{int}}(k) = \sum_m E(k,|m|) ,
\label{EK}
\end{equation}
and
\begin{equation}
\overline{E}_{\mathrm{int}}(|m|) = \sum_k E(k,|m|) .
\label{EM}
\end{equation}
Cross-sectional spectra, $E_m(k)$ and $E_k(|m|)$,
are obtained from the two-dimensional energy spectra
as a function of horizontal wavenumbers $k$
along a certain density wavenumber $m$
and
as a function of density wavenumbers $m$
along a certain horizontal wavenumber $k$, respectively.

As well as in $(k, m)$ space,
we numerically obtain energy spectra in $(\omega, m)$ space.
It is made in the similar way how observational spectra are.
Namely, we choose 
a point on the ``surface'' of our numerical ocean. We then conduct a ``vertical mooring,'' recording a time series 
of the horizontal velocity $\bm{u}(\bm{x}_0, \rho; t)$ at a
 fixed horizontal position ${\bm x}_0$.
The kinetic energy spectra is defined as
\begin{eqnarray}
 K(\omega, m) = \frac{1}{2} |\widetilde{\bm{u}}(\bm{x}_0, m; \omega)|^2,
\end{eqnarray}
where $\widetilde{\bm{u}}(\bm{x}_0, m; \omega)$ is the Fourier component of the horizontal velocity with respect to the vertical and time series.
The integrated kinetic energy spectra $\overline{K}_{\mathrm{int}}(\omega)$ and $\overline{K}_{\mathrm{int}}(|m|)$,
and the cross-sectional kinetic energy spectra $K_m(\omega)$ and $K_{\omega}(|m|)$
are defined from the two-dimensional kinetic energy spectrum, $K(\omega, m)$,
in the same way as the energy spectra in $(k, m)$ space.

As explained in the introduction,
in the 1970's
only
one-dimensional spectra were widely available with the exception of
the IWEX experiment. 
To measure  $\overline{E}_{\mathrm{vertical}}(m)$ 
spectrum in the ocean, one can use vertical profilers at a given position.
To measure $\overline{E}_{\mathrm{time}}(\omega)$ 
spectrum, one calculates time series of the mooring current meters.
Garrett and Munk assumed that the spectrum is 
separable in $(\omega,m)$ space. In other words, the spectrum is 
the product of function of $\omega$ alone and function of $m$ alone. 
In functional form, this statement is written as Eq.~(\ref{Separable}).

Then $\overline{E}_{\mathrm{time}}(\omega)$ and $\overline{E}_{\mathrm{vertical}}(m)$
were properly normalized to reproduce the characteristic total energy density of 
internal waves. Then these functions were  chosen so that 
the resulting $(\omega,m)$ spectrum is consistent with both $\omega$ spectrum and $m$ spectrum. 
In particular, moored spectrum was chosen to be
\begin{equation}
\overline{E}_{\mathrm{time}} (\omega) \propto \frac{1}{\omega \sqrt{\omega^2-f^2} }  \, .
\label{TimeGM}
\end{equation}
Notice that it has an integrable peak at the inertial frequency $f$. Moreover it has  $\omega^{-2}$ dependence for high frequencies, as is prominently displayed in moored observations. 
Then the $m$ spectrum was chosen as 
\begin{eqnarray}
\overline{E}_{\mathrm{vertical}}(m) \propto
\frac{1}{m^2+{m^{\ast}}^2}.
\label{eq:GMvert}
\end{eqnarray}
Here $m^{\ast}$ is the characteristic wavenumber determined by scale height.
Detailed analysis of these choices will be presented in \citet{iwobsPL}. 

The assumption of separability (\ref{Separable}) 
 allows one to construct a function of two arguments, 
$E(\omega,m)$ out 
of two one-dimensional functions.
The resulting GM spectrum in $(\omega, m)$ space is given by
\begin{eqnarray}
 E_{\mathrm{GM}}(\omega, m) \propto \frac{1}{\omega \sqrt{\omega^2 - f^2}}
  \frac{1}{m^2+{m^{\ast}}^2}.
\label{GM}
\end{eqnarray}

However, if one relaxes the assumption of the separability (\ref{Separable}), 
it is more
than one way to obtain the two-dimensional spectrum, $E(\omega, m)$, to fit the observed
one-dimensional spectra,
Eqs.~(\ref{TimeGM}) and (\ref{eq:GMvert}).
Moreover, it recently became
quite apparent that the separability (\ref{Separable}) is not satisfied
in the oceans \cite{kurtPC}. Therefore, we could argue that reality and 
accuracy of the separability (\ref{Separable}) perhaps deserves a closer examination. 
Furthermore, it may be possible that the one-dimensional spectra 
admits explanation without the separability.
We therefore stress through the paper the conceptual difference between the
integrated $\overline{E}_{\mathrm{int}}(\omega)$ and $\overline{E}_{\mathrm{int}}(m)$ spectra that could be confirmed by
oceanographic measurements and the two-dimensional spectra which is a
function of both $\omega$ and $m$.

A detailed analysis of the 
observational data is outside of the scope of the present paper. We will 
present such analysis in  \citet{iwobsPL}. 
There  details of IWEX and other experiments will be presented and 
reanalyzed. We will also present there results and 
analysis of more modern observations.
Brief catalog of historically available observational programs in the 
past three decades, along with characterization of deviations from 
the GM spectrum, is available in \citet{lvov-2004-92}.

The linear dispersion relation (\ref{eq:dispersion}) links
$\omega$ and $m$ with the horizontal wavenumber $k$. With the linear
dispersion relation, the spectrum can be transformed from both wavenumber
space, $(k,m)$, into frequency--horizontal-wavenumber space, $(k,\omega)$,
or frequency--vertical-wavenumber space, $(m,\omega)$. 
In particular,
the GM spectrum in $(k, m)$ space is given as
\begin{eqnarray}
\displaystyle
E_{\mathrm{GM}}(k,m) &=&
 E_{\mathrm{GM}}(\omega, m) \frac{\partial \omega}{\partial k}
\nonumber\\
 &\propto&
 \frac{1}{f^2 + \frac{g^2}{\rho_0^2 N_0^2} \frac{|\bm{k}|^2}{m^2}}
 \frac{1}{|m| \left(m^2+{m^{\ast}}^2\right)}
.\label{GMkm}
\end{eqnarray}
By using Eq.~(\ref{GMkm}) we can define the integrated $m$ spectra,
\begin{eqnarray}
 \overline{E}_{\mathrm{int,GM}}(m) &=& \int_{f}^{N_0} E_{\mathrm{GM}}(\omega, m) d\omega
\nonumber\\
  &=& \int_0^{\frac{\rho_0^2 N_0^2}{g^2} |m| \sqrt{N_0^2 - f^2}} E_{\mathrm{GM}}(k,m) dk.
\end{eqnarray}

In the large-wavenumber and large-frequency limit,
Eq.~(\ref{GM}) has the 
self-similar asymptotic form given by 
\begin{equation}
E_{\mathrm{GM}}(\omega,m) \propto \omega^{-2} |m|^{-2}.
\label{GMBigOM}
\end{equation}
Similarly, in the large-wavenumber limit, Eq.~(\ref{GMkm}) has a form 
\begin{equation}
E_{\mathrm{GM}}(k,m) \propto k^{-2} |m|^{-1} .
\label{GMBigkm}
\end{equation}
Both Eqs.~(\ref{GMBigOM}) and (\ref{GMBigkm}) should give the same one-dimensional spectrum as a function of vertical wavenumbers,
\begin{equation}\overline{E}_{\mathrm{int,GM}}(m) \propto |m|^{-2}.
\label{GMBigM}
\end{equation}
The difference between the exponents of $m$ in Eqs.~(\ref{GMBigkm}) and (\ref{GMBigM})
comes from the non-separability of Eq.~(\ref{GMkm}) in the $(k,m)$ space.

Note that the power-law exponents of one-dimensional spectra and those of two-dimensional spectra do not always coincide.
This could be explained by the fact that 
the exponents of the integrated spectra are a {\em weighted\/} mean
of those of the cross-sectional spectra, see (\ref{EK}) and (\ref{EM}). 
Therefore, the exponents of the integrated spectra can be different from those of the cross-sectional spectra
unless the spectra are separable.

\section{Numerical results\label{Results}}

\subsection{Run I}

\begin{figure}
 \begin{center}
  \includegraphics[scale=0.8]{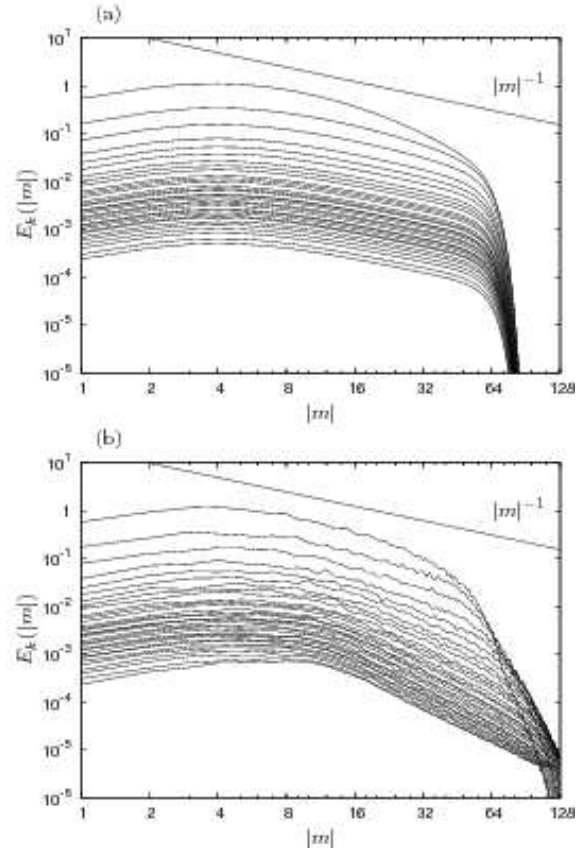}
 \end{center}
 \caption{Cross-sectional spectra $E_k(|m|)$
 of GM spectrum as the initial condition (upper)
 and
 of energy spectrum after about 35 hours (bottom)
of Run I.
 Significant differences in the region $16 < |m| < 64$
 indicates that GM spectrum is not statistically steady.
 }
\label{fig:gm_ns}
\end{figure}

Since the GM spectrum was believed to be the universal oceanographic spectrum,
the natural question to ask is whether the GM spectrum could be a
steady state of our numerical wave model. We realize that such a guess may be 
far-fetched, but it is nevertheless appealing to check it. 
Therefore we are led to examine
statistical stability of the GM spectrum. We assume that the steady
state of the ocean can be characterized by an energy flux from pumping regions
to damping regions in the wavenumber space.  Therefore to
achieve a truly steady state we need to model both pumping processes and
damping processes. As explained above, in modeling the damping we use
traditional hyper-viscosity approach in Eq.~(\ref{hyper}).
As we will see below, the resulting spectrum
depends upon the large-scale flows.
Therefore, the only way to see whether the GM spectrum is
close to being a steady state that is independent of the form of the
pumping is to choose no pumping at all. Thus we are led towards modeling the
system
as a freely decaying system, i.e. without external
forcing.

The GM spectrum with the cut-off in large horizontal and
density wavenumbers is employed as the initial energy spectrum.  The
cut-off is introduced to avoid decreasing accuracy of the
pseudo-spectral method.  The initial phases of complex amplitude,
$a(\bm{p})$, are given by uniformly distributed random numbers in $[0,2\pi)$.

Figure~\ref{fig:gm_ns} shows the initial spectrum and the
energy spectrum after about 35 hours in ocean time.  If the GM
spectrum were to be a universal steady state of our numerical model,
it would change very little in this simulation.  Instead, the density exponents rapidly
change from $-1$ to $-2$ only after $1.5$ days.  This spectral change
occurs faster than dissipation effects with timescale estimated to be
about 50 days at $|m|=32$.  Therefore it appears that the GM spectrum
is not a stable universal spectrum in the wavenumber region $16 < |m|
< 64$ at least for this particular numerical experiment.  The behavior 
of numerical experiment appears to
contradict the
arguments of  \cite{mccomas-1977-7,mccomas-1981-11} based on induced diffusion approach that all 
energy spectra are rapidly relaxed to the GM spectrum,
but is consistent with \cite{winters-1997}.

\subsection{Run II and III}\label{ssec:23}

We certainly realize that our numerical model of Run I may not fully describe
the ocean.
Then we question what could
be the nature of the statistical steady state of our
wave model of stratified rotating turbulence.
Here we model the pumping
phenomenologically. We assume that waves are forced at horizontally and vertically large scales, i.e. small wavenumbers. 
In particular we choose the forced wavenumbers such that 
\begin{eqnarray}
|\bm{k}_{\mathrm{F}}|^2 + |m_{\mathrm{F}}|^2\le 6^2. 
\label{eq:forcing23}
\end{eqnarray}
The forcing is modeled by fixing the amplitude of the forced wavenumbers to be constant in time. 
We also added an additional external dissipation $-D_{\omega} (\omega - N_0)^2$ for $\omega > N_0$ in these runs to avoid violation of the hydrostatic balance.
Detailed investigation of the non-hydrostatic effects is outside of the scope of present paper. 

Since the energy flows in the wavenumber space strongly depend on the inertial frequencies $f$ \cite{furuich-2005},
we performed Run II and III with differing values of the inertial frequencies $f$ 
(see Table~\ref{table:parameters}).  For Run II, the
inertial frequency $f=0.25 \times 10^{-4}$ rad/sec, while for Run III
the value of the inertial frequency is $f=1 \times 10^{-4}$ rad/sec. These
inertial frequencies correspond to latitudes of 10 degrees and 45 degrees.

Note that there is a ``critical'' latitude 
where the frequency of the semidiurnal tide is equal to twice the inertial frequency.
The principal difference
between these two runs is that Run II is southwards of 
``critical'' latitude in the northern hemisphere while Run III is northwards.

Our phenomenological forcing~(\ref{eq:forcing23}) corresponds to the band of frequencies greater than $4f$ for Run II
and $\sqrt{2} f$ for Run III. These values are the result of the discreteness of the numerical grid. 
Certainly the forcing is not necessarily characteristic of the ocean and is purely phenomenological, and chosen for numerical simplicity and to allow easy interpretation.

\begin{figure}
 \begin{center}
  \includegraphics[scale=0.8]{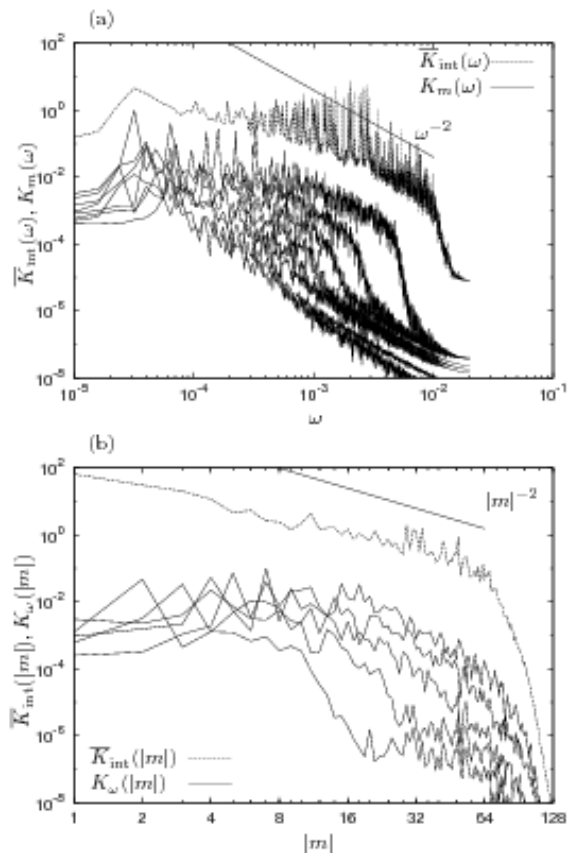}
 \end{center}
 \caption{
 Integrated kinetic energy spectra $\overline{K}_{\mathrm{int}}(\omega)$ and $\overline{K}_{\mathrm{int}}(|m|)$,
 and cross-sectional energy spectra $K_m(\omega)$ and $K_{\omega}(|m|)$
 of Run II.
 The kinetic energy spectra are obtained as functions of $\omega$ and $m$
 from time and vertical series of the horizontal velocity $\bm{u}(\bm{x}_0, \rho; t)$.
The GM spectrum scales as $\omega^{-2} |m|^{-2}$ for large frequencies and density wavenumbers.
The cross-sectional spectra that are functions of frequencies (left) are shown every eight curves for visibility.
The cross-sectional spectra that are functions of vertical wavenumbers (right) are shown 
when $\omega = 2, 4, 8, 16, 32 \times 10^{-4}$rad/sec.
}
\label{fig:sp_om_weak}
\end{figure}
\begin{figure}
 \begin{center}
  \includegraphics[scale=0.8]{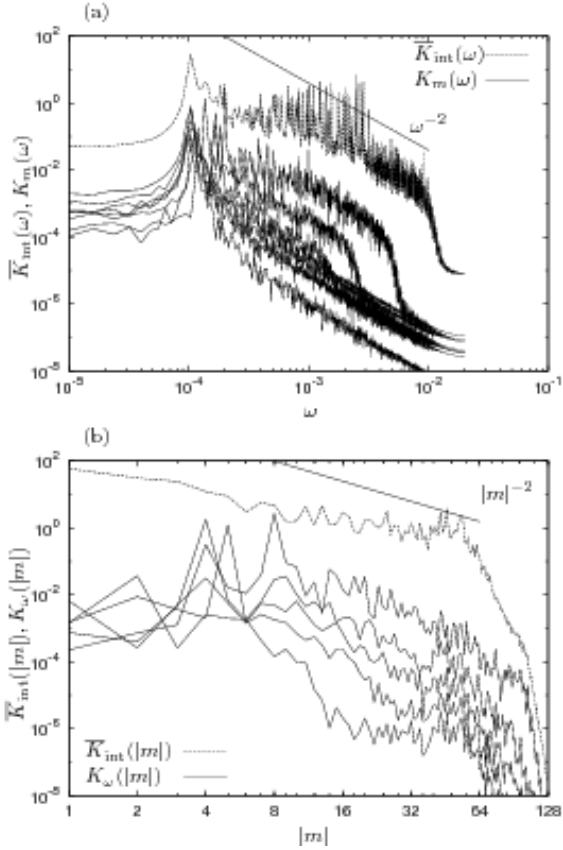}
 \end{center}
 \caption{
 Integrated kinetic energy spectra,
 and cross-sectional energy spectra
 of Run III.
 See the caption of Fig. \ref{fig:sp_om_weak} for details.
}
\label{fig:sp_om_mod}
\end{figure}

The kinetic energy spectra as functions of $\omega$ and $m$ of Run II and III
are shown in Figs.~\ref{fig:sp_om_weak} and \ref{fig:sp_om_mod}.
Both frequency spectra (left in Figs.~\ref{fig:sp_om_weak} and \ref{fig:sp_om_mod})
have peaks at the inertial frequencies.
This indicates that 
most energy accumulates in the near-inertial frequencies.
The behavior is characteristic of the ocean. Indeed,
these peaks correspond to the integrable singularity at the inertial frequency
of the GM spectrum (\ref{GM}).

After about $10^3$ days from the initial time
when all the wavenumbers have extremely small energy,
the systems are still transient
and the accumulation of energy in the near-inertial frequencies has not reached
complete statistically steady states.  
Thus the timescale of the development of the accumulation is relatively large.
Consequently one may conjecture that other processes, not present in our numerical model may affect the ocean at the large timescales. 
The most important of such processes which is not included in our simulations 
are the $\beta$ effects. Indeed, inclusion of $\beta$ effects would lead to 
existence of Rossby waves in our simulation, 
which may alter our results significantly. 
Other effects not present in our simulations,
which can affect larger scales,
may also alter the results.
However, our subject is
not to see how the energy spectra in the large-scale flows are formed
but rather to investigate how the energy spectra in the inertial subrange are formed.
For this subject,
our numerical model is consistent with observationally-based intuition 
that wave-wave interactions is the predominant mechanism of forming  
the internal-wave spectrum in the inertial subrange.

The energy spectrum of Run II has only weak accumulation of energy in the near-inertial frequency.
This could be qualitatively explained by the fact that all the forced wavenumbers have frequencies greater 
than $4f$ for Run II.
Consequently, PSI is effective in transferring energy to frequencies around $2f$ and large vertical wavenumbers.
Then PSI can no longer be effective in transferring energy towards the accumulation of energy in the near-inertial frequencies. Indeed, second 
PSI transfer would make the vertical wavenumber much larger and thus would move away from the accumulation. 
On the contrary, 
the energy spectrum of the Run III has moderate accumulation,
stronger than that of Run II.
This could be explained by the fact that the forced modes have frequencies greater than $\sqrt{2} f$.
Then PSI
can be effective in 
transferring energy from the wavenumbers that have frequency $2f$ to the accumulation whose frequencies are close to $f$
\cite{furuich-2005}.

Since the forcing does not have just one frequency but several frequencies,
the integrated spectra $\overline{K}_{\mathrm{int}}(\omega)$ have
huge oscillations corresponding to the linear frequencies of the forced wavenumbers.
Therefore, the power-law region of the frequencies, $10^{-4} \lesssim \omega \lesssim 10^{-2}$,
are strongly contaminated by the forcing.
However, the cross-sectional spectra with $|m| > 6$ are not much affected by the forcing.

The vertical-wavenumber spectra
(right in Figs.~\ref{fig:sp_om_weak} and \ref{fig:sp_om_mod})
appear to exhibit neither separable nor self-similar patterns.
On the other hand,
the integrated spectra do exhibit  the self-similarity.
The integrated spectra of the vertical wavenumbers are made mainly from the peaks at the near-inertial frequencies.
Therefore, the cross-sectional spectra do not always have similarities to the integrated spectra.

\begin{figure}
 \begin{center}
  \includegraphics[scale=0.8]{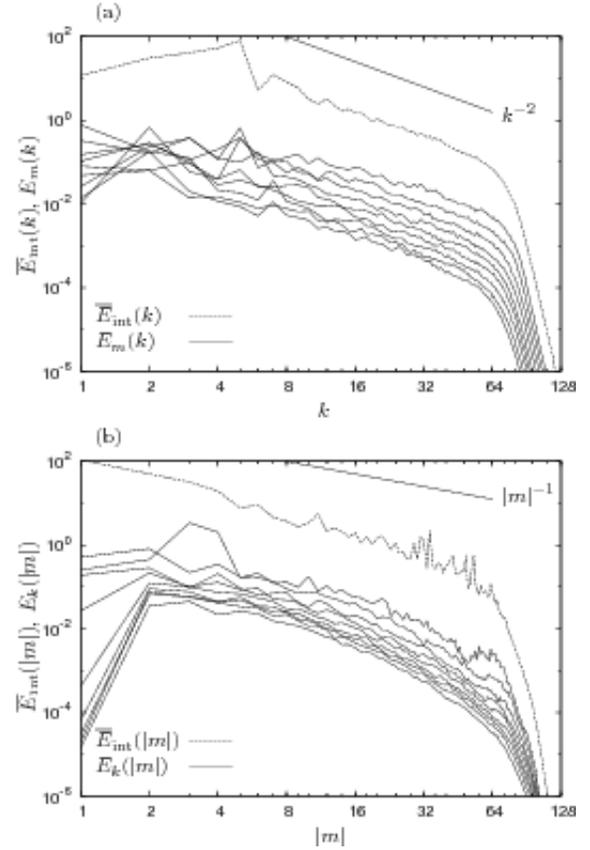}
 \end{center}
 \caption{
 Integrated energy spectra $\overline{E}_{\mathrm{int}}(k)$ and $\overline{E}_{\mathrm{int}}(|m|)$,
 and cross-sectional energy spectra $E_m(k)$ and $E_k(|m|)$
of Run II.
 The GM spectrum scales as $k^{-2} |m|^{-1}$ for large horizontal and density wavenumbers.
 The cross-sectional spectra are shown every four curves for visibility.
}
 \label{fig:energy_spectra_weak}
\end{figure}
\begin{figure}
 \begin{center}
  \includegraphics[scale=0.8]{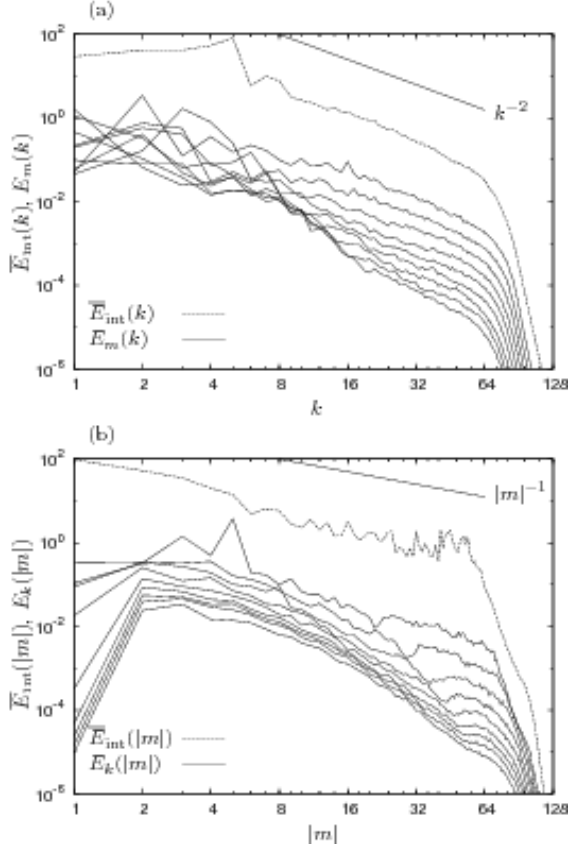}
\end{center}
 \caption{
 Integrated and cross-sectional spectra of Run III.
See the caption of Fig. \ref{fig:energy_spectra_weak} for details.
}
 \label{fig:energy_spectra_mod}
\end{figure}

It appears that it is advantageous to 
obtain $(k, m)$ spectra as well.
The integrated energy spectra $\overline{E}_{\mathrm{int}}(k)$ and $\overline{E}_{\mathrm{int}}(|m|)$,
 and cross-sectional energy spectra $E_m(k)$ and $E_k(|m|)$
of Run II and Run III are shown in Fig.~\ref{fig:energy_spectra_weak} and 
in Fig.~\ref{fig:energy_spectra_mod}, respectively.
The figures indicate that  the energy 
spectra as functions of $k$ and $m$
appear to be more separable than the kinetic energy spectra as functions of $\omega$ and $m$
in Run II and III.

\begin{figure}
 \begin{center}
  \includegraphics[scale=0.8]{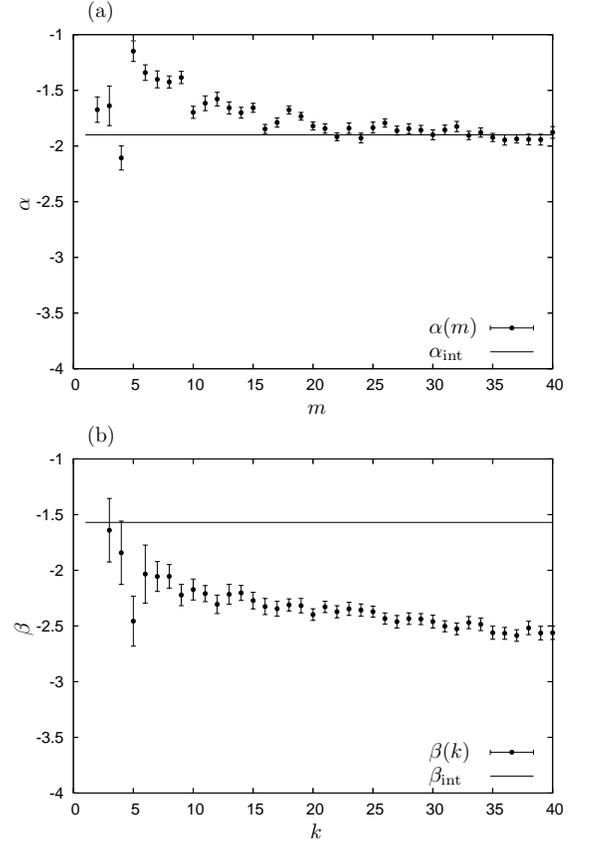}
\end{center}
  \caption{
 Power-law exponents of each cross-sectional spectrum in Run II.
 left: $\alpha(|m|)$, which are the exponents of the cross-sectional energy spectra $E_{m}(k)$,
 right: $\beta(k)$, which are the exponents of the cross-sectional energy spectra $E_{k}(|m|)$.
The error bars are obtained by fitting errors due to the least-square method.
 The exponents of the integrated spectra are also shown.
 }
  \label{fig:exp_weak}
\end{figure}
\begin{figure}
 \begin{center}
  \includegraphics[scale=0.8]{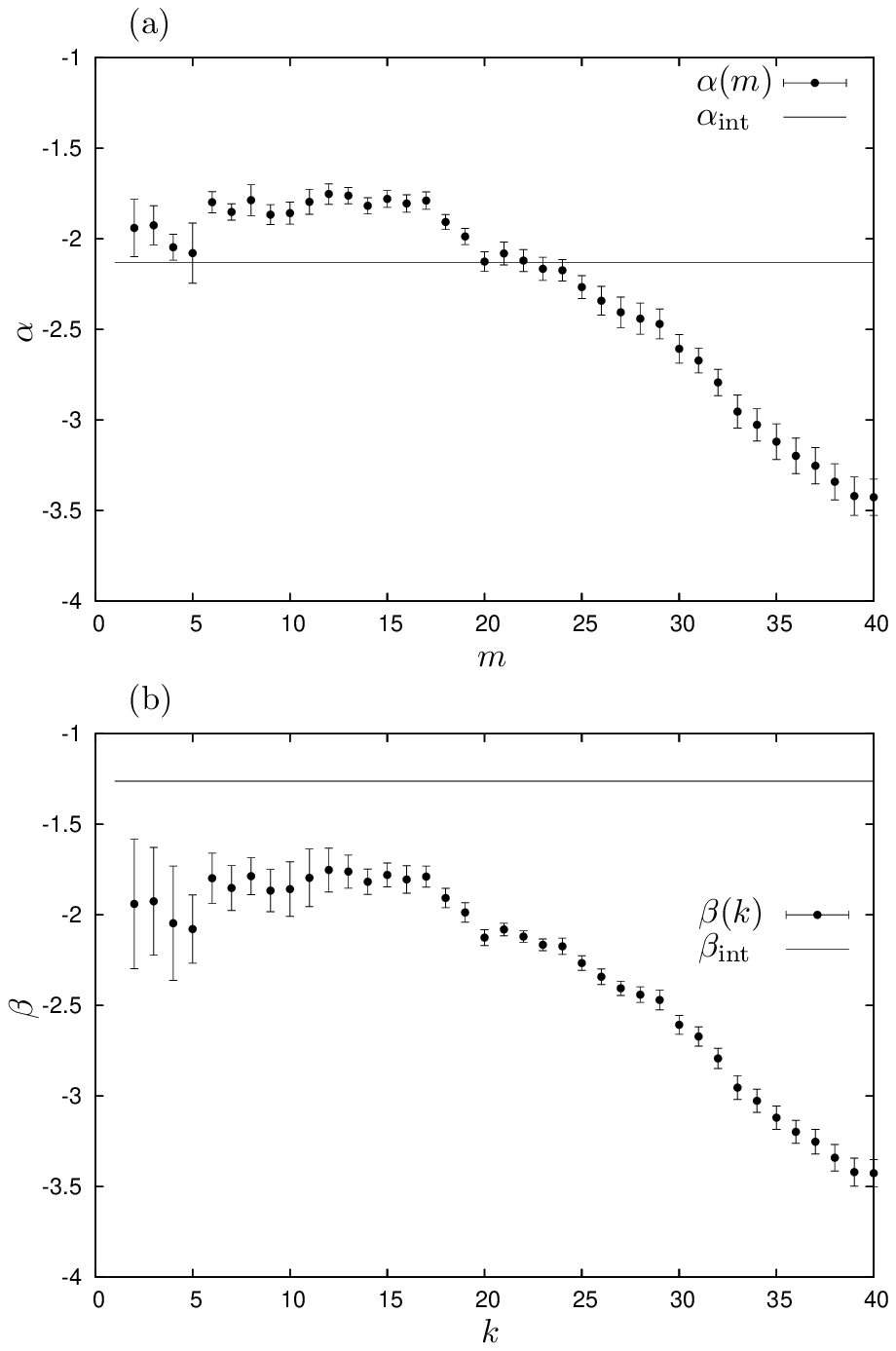}
 \end{center}
  \caption{
 Power-law exponents of Run III.
See the caption of Fig.~\ref{fig:exp_weak} for details.
 }
  \label{fig:exp_mod}
\end{figure}

The integrated and cross-sectional spectra in Figs.~\ref{fig:energy_spectra_weak} and \ref{fig:energy_spectra_mod}
have power-law regions.
The power-law exponents of the integrated and cross-sectional spectra are shown in Figs.~\ref{fig:exp_weak} and \ref{fig:exp_mod}.
The power-law exponents of the cross-sectional spectra by the least-square fitting
as
\begin{eqnarray*}
 E_{m}(k) \propto k^{\alpha(|m|)} \quad \mathrm{in} \ k \in [8,40]
\end{eqnarray*}
and
\begin{eqnarray*}
 E_{k}(m) \propto |m|^{\beta(k)} \quad \mathrm{in} \ |m| \in [8,40].
\end{eqnarray*}
The power-law exponents of the integrated spectra are also obtained
as
\begin{eqnarray*}
 \overline{E}_{\mathrm{int}}(k) \propto k^{\alpha_{\mathrm{int}}}
\end{eqnarray*}
and
\begin{eqnarray*}
 \overline{E}_{\mathrm{int}}(|m|) \propto |m|^{\beta_{\mathrm{int}}}
\end{eqnarray*}
in the same wavenumber regions.
In Run II
the energy spectrum is close to double-power 
\begin{equation}
E(k,m) \propto k^{-2} |m|^{-2.5}
 \label{SiamSpec1}
\end{equation}
in the large horizontal and density wavenumbers.

The exponents of the horizontal wavenumbers
of both integrated and cross-sectional spectra,
$\alpha_{\mathrm{int}}$ and $\alpha(m)$,
roughly agree with the those of the GM spectrum, $-2$ of
Eq.~(\ref{GMBigkm}). 
This fact supports that the oceanic spectra of the internal waves 
can be explained by the wave-wave interactions.
The integrated spectra are the sum of the cross-sectional spectra.
The integrated spectra of the horizontal wavenumbers in the inertial subrange
are not affected by the accumulation of energy around the horizontally longest waves. 

Indeed, to obtain the integrated spectra as functions of horizontal wavenumbers,
we use Eq.~(\ref{EK}) and integrate over all $m$ value for each $k$ value.
Thus,
the integrals over vertical wavenumbers in large horizontal wavenumbers are unaffected
by the strong presence of the near-inertial accumulation.
Therefore the integrated spectra of $k$ in large horizontal wavenumbers
well reflect the behavior of the spectra in the inertial subrange, and are immediately insensitive to the 
details of the accumulation of energy at small horizontal wavenumbers.
Similar arguments 
can be applied to the integrated spectra of frequencies.
Therefore, the integrated spectra of $\omega$ in high frequencies are determined by the behavior in the 
inertial subrange and are insensitive to the presence of the accumulation of energy 
at the near-inertial waves.

Consequently, the exponents of horizontal wavenumbers of both the integrated and cross-sectional spectra, $-2$,
are characteristic in our numerical experiments, 
and is consistent with oceanographic observations.

On the other hand, to obtain the integrated spectra as functions of vertical wavenumbers,
one should use
Eq.~(\ref{EM}) to integrate for each $m$ value over all $k$ values.
Therefore, when integrating over $k$ values, the value of the
integrals are always strongly affected by the accumulation of energy at small
$k$ values.  Therefore the exponents of the integrated spectra of $m$
are determined exclusive by the near-inertial accumulation, and are insensitive
to the behavior in the inertial subrange.  Consequently, the integrated spectra of $m$
are inconsistent in our simulations, and are in fact accumulation-dependent. 

In actual fact,
the integrated spectra as functions the vertical wavenumbers
are less steep than the GM spectrum.
Meanwhile, the cross-sectional spectra as functions the vertical wavenumbers
in the inertial subrange
are steeper than the GM spectrum.
It comes from the fact that
the exponents of the integrated spectra are a weighted mean
of those of the cross-sectional spectra. 
In other words, $\beta_{\mathrm{int}}$ strongly depends on $\beta(1)$ and $\beta(2)$.
Note that the horizontal wavenumbers $k=0$ corresponds to the inertial frequencies.

\begin{figure}
\begin{center}
 \includegraphics[scale=0.8]{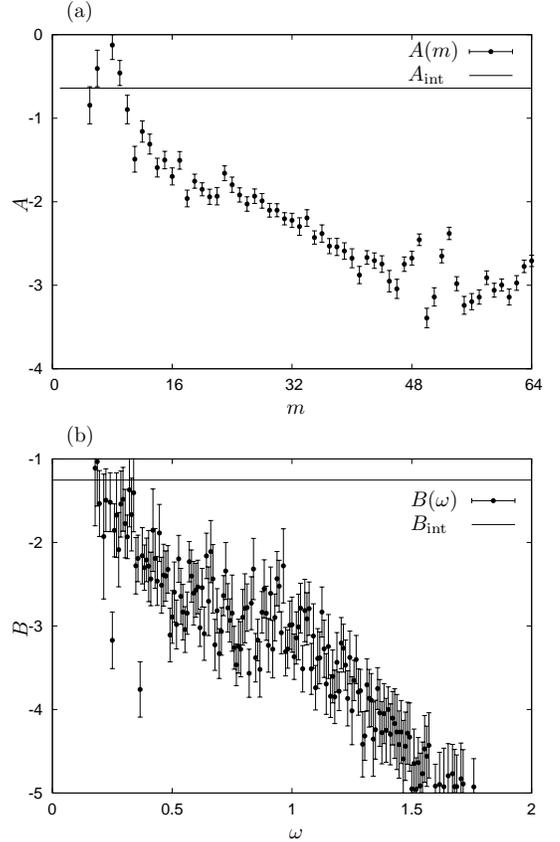}
\end{center}
 \caption{
 Power-law exponents of each cross-sectional spectrum of the kinetic energy in Run II.
 left: $A(|m|)$, which are the exponents of cross-sectional energy spectra $K_{m}(\omega)$ in Fig.~\ref{fig:sp_om_weak},
 right: $B(\omega)$, which are the exponents of cross-sectional energy spectra $K_{\omega}(|m|)$ in Fig.~\ref{fig:sp_om_weak}.
}
  \label{fig:exp_om_weak}
\end{figure}
\begin{figure}
\begin{center}  
 \includegraphics[scale=0.8]{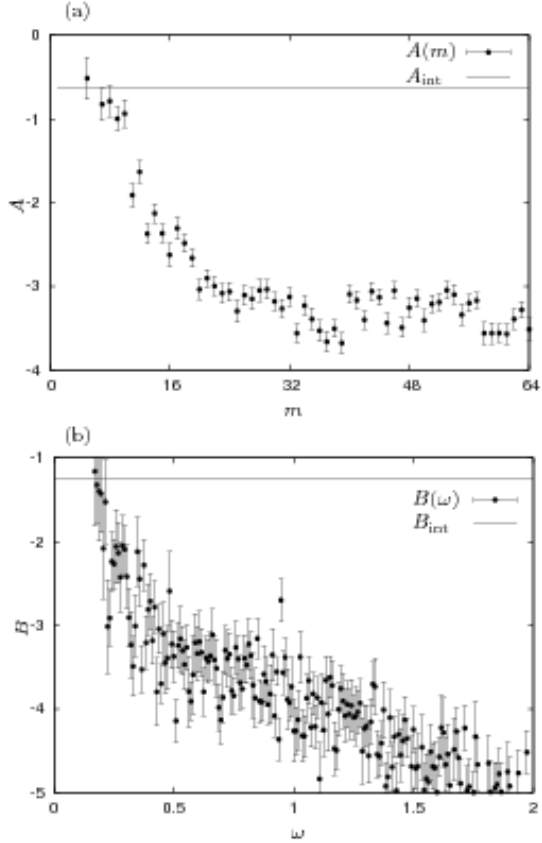}
\end{center}
\caption{
 Power-law exponents of each cross-sectional spectrum of the kinetic energy in Run III.
 See the caption of Fig.~\ref{fig:exp_om_weak} for details.
}
  \label{fig:exp_om_mod}
\end{figure}

The energy spectrum does not have double-power laws
in the large wavenumbers
in Run III.
It can be explained by contamination of the inertial subrange in the small horizontal wavenumbers and the large density wavenumbers,
where the cross-sectional spectra do not show self-similarity in Fig.~\ref{fig:energy_spectra_mod}.
It is curious that another double-power law
appears in Run III in the small horizontal and density wavenumbers.
Indeed, the spectrum at small wavenumbers can be approximated by
\begin{equation}
E(k,m) \propto k^{-2} |m|^{-2}.
\label{SiamSpec2}
\end{equation}
As this power law does not appear in the inertial subrange,
it has limited relevance to the scope of this paper.

Similar to the cross-sectional spectra in $(k,m)$ space,
the spectral exponents in $(\omega,m)$
space could also be measured. 
The power-law exponents of the integrated and cross-sectional spectra
of the kinetic energy in Figs.~\ref{fig:sp_om_weak} and \ref{fig:sp_om_mod}
are shown in Figs.~\ref{fig:exp_om_weak} and \ref{fig:exp_om_mod}.
The least-square fittings are made in $\omega \in [10^{-4},10^{-3}]$rad/sec and $|m| \in [8,32]$
as $K_{m}(\omega) \propto \omega^{A(|m|)}$ and
$K_{\omega}(m) \propto m^{B(|\omega|)}$.
We cannot find any characteristic exponents in the figures.
Therefore it appears that the spectral energy density of these runs 
is not separable in the $(\omega,m)$ space.
This statement appears to be consistent with the fact that the cross-sectional spectra of the kinetic energy
are neither self-similar nor separable.

To summarize, Runs II and III exhibit the accumulation of energy at the near-inertial 
waves.
The formation mechanism of the near-inertial accumulation
is consistent with the oceanographic PSI arguments.
Furthermore, 
the cross-sectional spectra in Figs.~\ref{fig:energy_spectra_weak} and \ref{fig:energy_spectra_mod}
appear to be more separable than those in Figs.~\ref{fig:sp_om_weak} and \ref{fig:sp_om_mod}.
It is supported by the fact that
constant-exponent regions appear
in the spectra in $(k ,m)$ space
(Figs.~\ref{fig:exp_weak} and \ref{fig:exp_mod})
but not in in the spectra in $(\omega ,m)$ space
(Figs.~\ref{fig:exp_om_weak} and \ref{fig:exp_om_mod}).
Therefore,
the spectrum appear to be more separable in $(k,m)$ space
than in $(\omega,m)$ space in Run II and III.
It also appears that the spectrum in the inertial subrange is sensitive 
to the details of the near-inertial accumulation.
Most important, we reproduce the $-2$ exponent of the integrated spectra of the horizontal wavenumbers.

\subsection{Run IV}
\begin{figure}
 \begin{center}
  \includegraphics[scale=0.65]{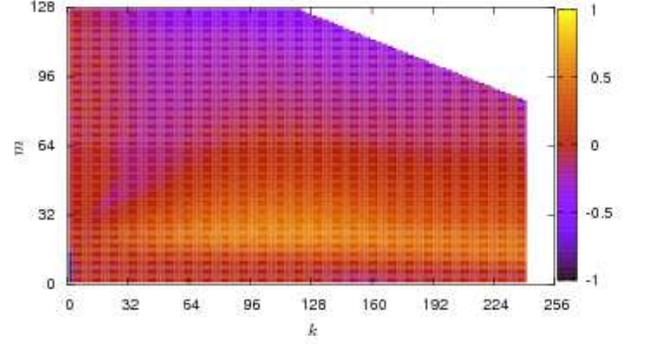}
 \end{center}
 \caption{Relative difference between energy spectra
 after about $1.5$ days
 developed from the GM spectrum with and without energy in small horizontal wavenumbers
 $E_{\mathrm{d}}(k,|m|)$.
 }
 \label{fig:gm_nl}
\end{figure}

To further illustrate the influence of the accumulation around the horizontally longest waves, and to investigate the importance of the nonlocal interactions in the wavenumber space,
we perform Run IV of Table~\ref{table:parameters}.
There we choose the initial condition
same as the Run I but with no energy in $k<3$ and $|m| < 16$.
The energy spectrum after about $1.5$ days is denoted by $E_{\mathrm{nl}}(k,|m|)$.
The energy spectrum developed from the GM spectrum in Run I that is shown in Fig.~\ref{fig:gm_ns}(right)
is denoted by $E_{\mathrm{GM}}(k,|m|)$.
Figure~\ref{fig:gm_nl} shows the relative difference defined as
\begin{eqnarray}
E_{\mathrm{d}}(k,|m|) = \frac{E_{\mathrm{nl}}(k,|m|) - E_{\mathrm{GM}}(k,|m|)}{E_{\mathrm{GM}}(k,|m|)} \, .
\end{eqnarray}
The wavenumbers different in the initial conditions appear as a black rectangle in bottom left in Fig.~\ref{fig:gm_nl}.
If the nonlocal interactions with the accumulation of energy around the horizontally longest waves were not dominant,
the relative difference in the inertial subrange in Fig.~\ref{fig:gm_nl} would be small
or slightly negative
since the nonlinear interactions try to compensate the defect in the small wavenumbers.
Instead, $E_{\mathrm{nl}}(k,|m|)$
has more than 20\% energy in $10 \lesssim |m| \lesssim 40$
and less energy transfer to the dissipation region in $|m| \gtrsim 80$.
This suggests that
the small wavenumbers in the accumulation of energy
transfer energy
from small density wavenumbers to large density wavenumbers
in the inertial subrange.
This scenario of the energy transfer to the large density wavenumbers is 
qualitatively consistent with
the Induced Diffusion mechanism \cite{mccomas-1977-7,mccomas-1981-11}.

\subsection{Run V}
Run II and III have indicated that the external forcing at multiple
frequencies does not produce smooth spectral energy density in
the frequency space.
We therefore are led to the question what would happen if the forced wavenumbers correspond to the single frequency. 
In Run V,
the forced wavenumbers are modeled as M$_2$ tides
and have an almost single frequency of $3f$. The single frequency forcing 
is significantly different from the band of forced frequencies in Run II and III.
One of the reasons to
choose $3f$ is to further confirm that PSI mechanism is effective.
Indeed, PSI is dominant in transferring energy to small $k$ and large $m$ wavenumbers
when the forced frequencies are greater than $2f$.
Consequently, with $3f$ forcing PSI will create a second peak at $1.5 f$. 
Although one could choose $2f$ as a forced frequency,
we have chosen $3 f$ in order to be able to distinguish 
the direct excitations that appear at $1.5f$ via PSI
and the accumulation of energy at the near-inertial frequencies around $f$.

Fixing forced frequency to be $3 f$, we still have a freedom of choosing forced wavenumbers. 
For simplicity, 
to implement the forcing numerically,
the external forcing is added to 24 small wavenumbers whose frequencies are close to $3f$.
Specifically, we keep the amplitudes of canonical variables, $a(\bm{p})$,
to be  constant in time for the following wavenumbers:
$(\bm{k}, m) = (\pm 1, 0, \pm 2)$, $(0, \pm 1, \pm 2)$, $(\pm 1, \pm 1, \pm 3)$,
$(\pm 2, 0, \pm 4)$ and $(0, \pm 2, \pm 4)$.

\begin{figure}
 \begin{center}
  \includegraphics[scale=0.65]{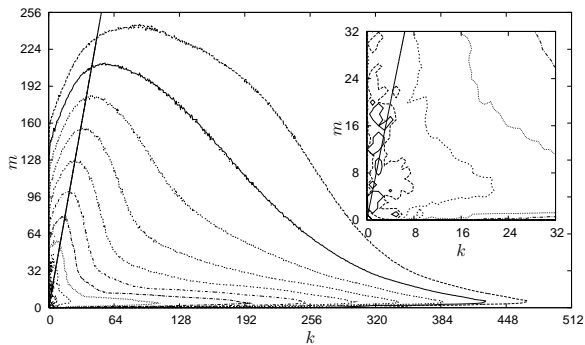}
 \end{center}
 \caption{Two-dimensional energy spectrum $E(k,|m|)$ of Run V.
 The contours are plotted every powers of ten from $10^{-10}$ to $1$.
 The straight line from the origin shows the wavenumbers $\omega=1.5f$.
 The inset is the enlargement of the area near the origin.
 }
 \label{fig:energy_spectrum_2D}
\end{figure}
 
In Fig.~\ref{fig:energy_spectrum_2D}
we show the two-dimensional energy spectrum close to a statistically steady state
obtained in this simulation.
We observe that there is a significant energy accumulation around $k \sim 1$,
i.e. the horizontally longest waves. 
The wavenumbers that have $1 \leq  k \leq 2$ except the forced wavenumbers,
which are considered as the accumulation of energy,
account for approximately 60\% of the total energy.
Appearance of the accumulation in the near-inertial waves is consistent with 
oceanographic observations. It is also consistent with Runs II and III. 
This consistency in observing the accumulation of energy in the near-inertial waves
is one of the main results of the present paper. 

\begin{figure}
 \begin{center}
  \includegraphics[scale=0.8]{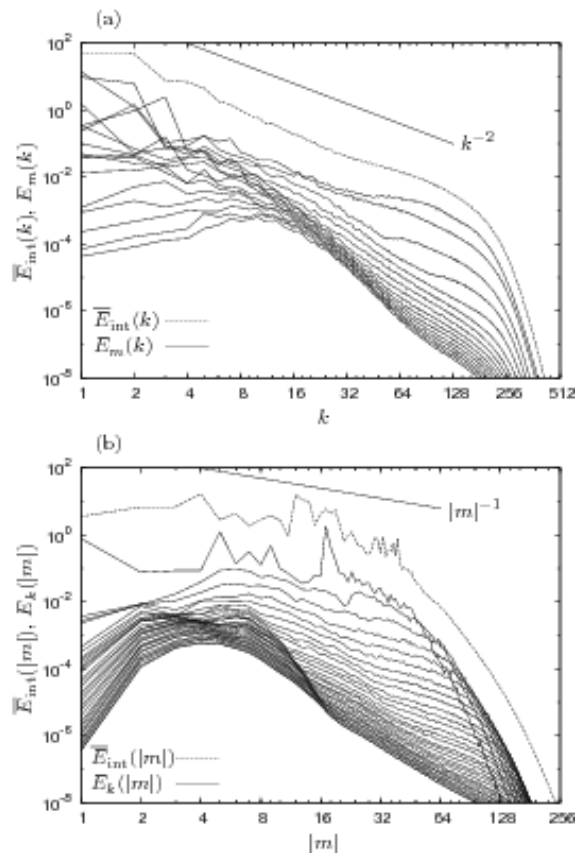}
 \end{center}
 \caption{
 Integrated and cross-sectional spectra of Run V.
See the caption of Fig. \ref{fig:energy_spectra_weak} for details.
 }
 \label{fig:spectra}
\end{figure}

\begin{figure}
 \begin{center}
  \includegraphics[scale=0.8]{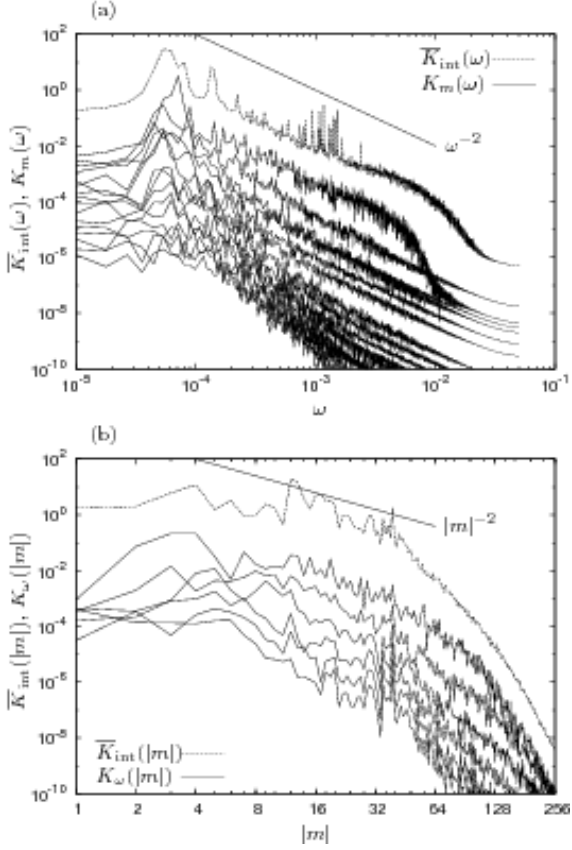}
 \end{center}
 \caption{
 Integrated kinetic energy spectra,
 and cross-sectional energy spectra
 of Run V.
The cross-sectional spectra that are functions of frequencies (left) are shown every eight curves for visibility.
The cross-sectional spectra that are functions of vertical wavenumbers (right) are shown 
when $\omega = 2.5, 5, 10, 20, 40 \times 10^{-4}$rad/sec.
 See the caption of Fig.~\ref{fig:sp_om_weak} for details.
}
\label{fig:om}
\end{figure}

Aside from the accumulation of energy around the horizontally longest waves,
other excited wavenumbers appear around the line $\omega = 1.5 f$.
The frequency is half of the frequency of the forced wavenumbers.
This excitation can be qualitatively be explained by the PSI mechanism. 
Indeed, the excited wavenumbers whose frequencies are close to $1.5 f$
make resonant triads with the forced wavenumbers
via PSI.
Furthermore, another excitation can be seen around $m \simeq 5$.
The excitation can be interpreted as resonant interactions with the accumulation of energy due to the ES mechanism.

Figure~\ref{fig:spectra} shows integrated spectra,
$\overline{E}_{\mathrm{int}}(k)$ and $\overline{E}_{\mathrm{int}}(|m|)$,
and cross-sectional spectra, $E_m(k)$ and $E_k(|m|)$,
obtained from the two-dimensional energy spectrum shown in Fig.~\ref{fig:energy_spectrum_2D}.
The integrated spectra can be interpreted roughly to be 
$$\overline{E}_{\mathrm{int}}(k) \propto k^{-1.98 \pm 0.02}$$
and
$$\overline{E}_{\mathrm{int}}(|m|) \propto |m|^{-1.33 \pm 0.31}.$$
The exponents are obtained with the least-square method
in the intervals of $k \in [8,128]$ for $\overline{E}_{\mathrm{int}}(k)$
and that of $|m| \in [4,32]$ for $\overline{E}_{\mathrm{int}}(|m|)$.
Note that the exponents are not too far from the large-wavenumber self-similar form of the GM spectrum.
Indeed, the large wavenumber asymptotic form of the GM spectrum is given by
Eq.~(\ref{GMBigkm}). 

We emphasize nevertheless that similarity of the integrated numerical
spectrum as a function of vertical wavenumbers with Eq.~(\ref{GMBigkm}) may be
coincidental to some degree.
The exponent of the integrated spectrum of the GM spectrum in large vertical wavenumbers is not $-1$ but $-2$.
In addition,
as apparent from Fig.~\ref{fig:spectra}, the spectrum in $(k,m)$ space
is not separable. Moreover, as explained above, the exponent of the vertical wavenumbers is
mostly determined by the accumulation of energy at small $k$. Consequently it 
is rather 
sensitive to the specifics of numerical experiments. On the other hand, the 
$k$ exponent is relatively insensitive to the accumulation and is determined by the inertial subrange. The behavior was also observed in Run II and III.

\begin{figure}
 \begin{center}
  \includegraphics[scale=0.8]{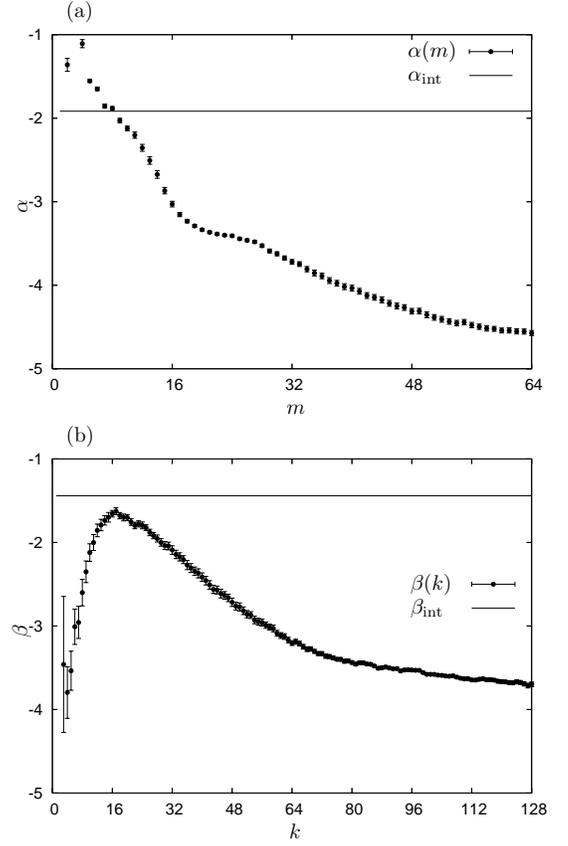}
 \end{center}
 \caption{
 Power-law exponents of Run V.
See the caption of Fig. \ref{fig:exp_weak} for details.
 }
 \label{fig:exp_css}
\end{figure}

\begin{figure}
 \begin{center}
  \includegraphics[scale=0.8]{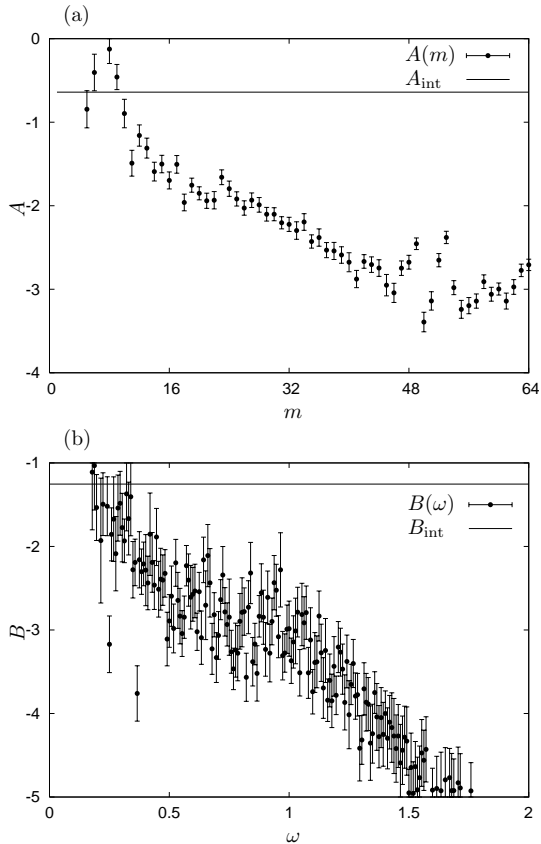}
 \end{center}
\caption{
 Power-law exponents of the kinetic energy in Run V.
 See the caption of Fig. \ref{fig:exp_om_weak} for details.
}
\label{fig:exp_om}
\end{figure}

We also measure the kinetic energy spectrum in $(\omega,m)$ space. 
Figure~\ref{fig:om} shows the results for Run V.
It appears that some energy exists in the large frequencies greater than the buoyancy frequency. It comes mainly from the non-periodicity of the time series, and is called the Gibbs phenomenon.

Observe that the integrated spectrum of frequency
is fitted by 
$$\overline{E}_{\mathrm{int}}(\omega) \propto \omega^{-2.19 \pm 0.11}.$$
over $\omega \in [10^{-4}, 10^{-3}]$. The asymptotic behavior is consistent
with the GM spectrum and oceanographic observations. 
This is the main result of this paper. 
Also note that the vertical spectrum could be fitted by $|m|^{-1.12 \pm 0.35}$
over $|m| \in [4,32]$. 
 
Furthermore, one can interpret Fig.~\ref{fig:om} as having three peaks in $\overline{E}_{\mathrm{int}}(\omega)$.
The accumulation of energy makes
the peak around the inertial frequency $f$ of this run, namely $5 \times 10^{-5}$rad/sec.
The second peak can be found around $7 \times 10^{-5}$rad/sec
which is the half of the frequency of the forced wavenumbers.
Appearance of the excitation at half the forcing frequency is consistent with the PSI mechanism.
The third peak is seen around $1.4 \times 10^{-4}$rad/sec
which is the frequency of the forced wavenumbers.

To illustrate the spectral details in the inertial subrange,
we show the power-law exponents of each cross-sectional spectrum in Fig.~\ref{fig:exp_css},
$\alpha(|m|)$ and $\beta(k)$.
The exponents are obtained by fitting
$E_{m}(k) \propto k^{\alpha(|m|)}$ in $k \in [16,128]$
and $E_{k}(m) \propto |m|^{\beta(k)}$ in $|m| \in [16,64]$.
The exponents of the integrated spectra, $\alpha_{\mathrm{int}}=-1.98$ and $\beta_{\mathrm{int}}=-1.33$,
are also shown in Fig.~\ref{fig:exp_css}.
The main point taken from the figure is that
the integrated spectra have different exponents from the ones of the cross-sectional spectra in the inertial subrange.
We observe that $\alpha$ varies between $-1$ and $-5$. 
Similarly, $\beta$ varies between $-1.5$ and $-3.5$. 

Similar to Runs II and III,
the exponent of the integrated spectrum of the horizontal wavenumbers
is close to the GM spectrum.
However, the cross-sectional spectra are steeper than the GM spectrum
in the large horizontal and vertical wavenumbers considered as the inertial subrange.
This can be explained by the fact that they are {\em contaminated\/} by direct excitations due to PSI and ES mechanisms.
Apparently,
the cross-sectional exponents are not constant.
Therefore,
the energy spectrum of the numerical simulation cannot be accurately fitted by double-power functions.

As the integrated and cross-sectional spectra of the GM spectrum have different exponents with respect to the vertical wavenumbers,
the exponents of the integrated spectra
are much different those of the cross-sectional spectra in the inertial subrange
also in our simulations.
This discrepancy is caused by the non-separability of the two-dimensional spectrum.
The discrepancy in the vertical-wavenumber exponents is especially clear by the near-inertial accumulation.
The integrated spectrum, $\overline{E}_{\mathrm{int}}(m)$, is established
mainly by $E_{k=1,2}(m)$ i.e. the accumulation of energy.
Therefore,
the exponent of the integrated spectrum is determined exclusively
by the exponents of the cross-sectional spectra in the small horizontal wavenumbers. 

Figure \ref{fig:exp_om} shows
power-law exponents of each cross-sectional spectrum of the kinetic energy
as functions of $\omega$ and $|m|$.
The fits are made over $\omega \in [10^{-4}, 10^{-3}]$rad/sec or over $|m| \in [8,32]$.
The exponents of the integrated spectra are also shown in the figure for reference.
It is instructive to compare Figs.~\ref{fig:exp_css} and \ref{fig:exp_om}.
It appears that the $(\omega,m)$ spectrum is more separable than the $(k,m)$ spectrum. Indeed, 
the exponent $\alpha$ of horizontal wavenumbers of the $(k,m)$ spectrum of Fig.~\ref{fig:exp_css} varies between $-1$ and $-5$,
while the exponent $A$ of frequencies of the $(\omega,m)$ spectrum of Fig.~\ref{fig:exp_om}
varies between $-1.5$ and $-4$. Furthermore, it appears that $A$ can loosely be represented as having the value of $A \simeq -2.7$. 
Similarly, apart from the small-wavenumber and low-frequency region,
the exponent $\beta$ of vertical wavenumbers of the $(k,m)$ spectrum varies between $-2.5$ and $-3.5$,
and
the exponent $B$ of vertical wavenumbers of the $(\omega,m)$ spectrum can loosely be interpreted as 
having value around $-3.5$. However, the separability (\ref{Separable}) can not be interpreted as being fully satisfied.
Our observation of the spectrum of Run V being more separable in $(\omega,m)$ space than in $(k,m)$ space is consistent with observationally-based intuition. 
We again note that the exponents of the cross-sectional spectra are also far from those of the integrated spectra.
This statement is true for both $(k,m)$ and $(\omega,m)$ spectra.

As explained above,
the near-inertial accumulation appears only in the small horizontal wavenumbers
and the small frequencies.
Thus,
the accumulation does not affect the integrated spectrum of $k$ in the large horizontal wavenumbers
when the integration of the two-dimensional $(k,m)$ spectrum over $m$
to obtain the integrated spectrum is made.
Similarly,
the integrated spectrum of $\omega$ in high frequencies are not affected by the near-inertial accumulation.
Then,
we can argue that
the integrated spectra of the horizontal wavenumbers and the frequencies
are determined by the wavenumbers in the inertial subrange.
Consequently these integrated spectra appear to be characteristic in our 
numerical runs. 
Therefore,
the power-law exponents of the integrated spectra
roughly corresponding to those of the observations including the GM spectrum
reflects the nonlinear interactions of the internal-wave system.

To summarize, Run V is the largest and longest run that we have performed.
The internal-wave field
is forced by single frequency, small horizontal and vertical wavenumbers.
Run V clearly demonstrates the energy accumulation around the horizontally
longest waves. This run also shows that PSI mechanism is
effective at creating a second peak at half the forced
frequency. Furthermore, this run reproduce the integrated spectra,
$k^{-2}$ spectrum and $\omega^{-2}$ spectrum.
This is consistent with observations.
This run also demonstrates that
the spectrum in $(\omega,m)$ space is more
separable than in $(k,m)$ space. It is also
consistent with oceanic observations.

\section{Discussion}
\label{sec:discussion}

Stratified rotating turbulence is a complicated and fascinating subject. It has been a subject of intensive research in the last few
decades. In particular various numerical simulations were
performed.
For example, \citet{winters-1997} numerically modeled
stratified Navier--Stokes turbulence. They consider a depth dependent
buoyancy frequency, and focus on a depth dependence of various physical
quantities. Their simulation is done on a $32 \times 256 \times 129$ grid.
They implemented
realistic boundary conditions in the vertical directions
and obtained the energy dissipation rates consistent with theories and observations.
In contrast,
our numerical simulations focus on spectral energy density.

More recent simulations were also performed.
\citet{PhysRevE.68.036308} studied numerically stratified turbulence, and showed formation of pancake vortices
and considerable amount of energy in
the horizontally uniform vertical shear and vortical modes.
Numerical simulations were also performed by \citet{smith2002gsl} for rotating stratified turbulence and by \citet{smith2005nra} for rotating 
turbulence. These simulations 
demonstrate the tendency of energy to accumulate
at the horizontally large-scale flows.
They presented that the accumulation is caused mainly by resonant wave-wave interactions.
They also obtained one-dimensional spectra similar to our integrated spectra.

In contrast to most of these simulations,
our simulations are performed  in periodic boxes in all three directions.
Furthermore, we completely exclude vortices and horizontally uniform vertical shears.
In addition,
we use larger numerical grids, which are required
in order to analyze the inner structures of the energy spectra in the inertial subrange.
In addition, larger numerical grids are necessary to make the nonlocal interactions more pronounced in the simulations.

Certainly our reduced model constitutes a simplification of the
ocean.  One could argue 
that the horizontally uniform vertical shears excluded in our simulations
might play an important role for the wavenumbers in the inertial subrange.
Indeed,
the horizontally uniform shear might strongly affect the spectra in the inertial subrange
as the accumulation of energy at the horizontally longest waves does.
Furthermore, the largest-scale motions also depend on other effects 
which are excluded from our simulation. In particular, we excluded 
variability of the inertial frequencies or $\beta$ effect,
seasonal variability, and terrain properties.
They may significantly affect the accumulation, providing mechanisms to 
diminish or build the accumulation. Consequently the waves in the inertial subrange will be affected.  

As mentioned in Sec.~\ref{sec:introduction}, anisotropic wave turbulence
systems are often dominated by the nonlocal interactions.
The oceanic internal waves are not an exception.
Therefore it is important
to study the effects of the nonlocal interactions in the wavenumber space.

The invariance under Galilean transformation is broken in the systems 
which have dominance of the nonlocal interactions.
Indeed, Coriolis effect and presence of vertical stratification
breaks the Galilean invariance of the internal waves.
Consequently, one could argue that the dominance of the nonlocal
interactions is caused by violation of Galilean invariance of the systems.
In fact, in the systems where Galilean invariance is broken,
the large-scale flows not only advect the
small-scale flows but also actively exchange energy with them
almost like an inertia force.
It is in contrast with the typical behavior of
the large-scale motions in Galilean-invariant systems.
In Galilean-invariant systems
the large-scale flows advect small-scale flows,
as in the sweeping in Navier--Stokes turbulence.

\section{Summary}
\label{sec:summary}
In this work,
we performed numerical modeling of the reduced wave-only 
stratified rotating turbulence in order 
to investigate the wave-wave interactions.
In particular, we made a series of $5$ numerical runs with
different initial conditions, grid sizes and forcing. 

Run I is the freely decaying run with the GM spectrum employed as the initial condition. 
It has demonstrated that the GM spectrum is not a steady-state spectrum for our numerical model.

Runs II and III are forced--damped runs with different values of $f$. 
These runs consistently demonstrated the tendency of the energy 
to accumulate in the horizontally longest (i.e. near-inertial) waves.
In these runs the level of the accumulation depends on the relation between the
value of the inertial frequency $f$ and forced frequencies.
The accumulation of energy in the near-inertial frequencies is
characteristic of the oceans. 
The runs also 
demonstrated largely non-separable spectra.
It also appears that 
the energy spectrum in $(k,m)$ space is more separable than in $(\omega,m)$ space. This is at odds 
with oceanic intuition. We also observed realistic $k^{-2}$ spectrum in 
Run II. Furthermore, the formation of the inertial peak is qualitatively consistent with a ``named'' nonlocal interaction, which is the Parametric Subharmonic Instability.
It also appears that the integrated vertical-wavenumber spectrum is highly sensitive to the accumulation 
at the horizontally longest waves. Finally, 
the integrated frequency spectra cannot be fitted by power laws due to the forcing at multiple 
frequencies.

Run IV is the freely decaying run with the GM spectrum without the small wavenumbers as initial conditions. 
This run demonstrates that the internal-wave statistical
properties are strongly affected by the nonlocal interactions with the small
wavenumbers. In particular, removing the small wavenumbers from the
initial condition produces a significantly different outcome.
Thus, the energy transfer in the inertial subrange is qualitatively explained by the Induced Diffusion.

Run V also demonstrates that energy accumulates at
the horizontally longest near-inertial waves. This is consistent with 
Runs II and III as well as the ocean. 
It also appears that the two-dimensional
spectrum is more separable in $(\omega,m)$ space than in $(k,m)$
space. This is consistent with observationally-based intuition. Furthermore,
the integrated frequency spectrum of Run V is given by $\omega^{-2}$,
consistent with the ocean.
Similarly, the integrated horizontal-wavenumber spectrum $k^{-2}$ 
is also consistent with the observations.
However, the integrated vertical-wavenumber spectrum do not show the power-law exponent consistent with the observations.
This could be explained by the fact that 
the integrated vertical-wavenumber spectrum is determined by the accumulation at the near-inertial waves. The near-inertial 
waves may be much affected by processes not considered in this manuscript. 

In short, our numerical runs reproduce the following
behavior:
\begin{itemize}
\item Energy tends to accumulate at the horizontally longest waves i.e. near-inertial 
waves
\item To a lesser degree, some energy was observed in the small vertical wavenumbers.
\item Spectra in the inertial subrange are not completely separable
\item Realistic integrated horizontal $k^{-2}$ spectrum is observed
\item Realistic integrated frequency  $\omega^{-2}$ spectrum is observed
\item The integrated vertical-wavenumber spectrum is determined exclusively by the near-inertial 
waves
\item The spectrum in the inertial subrange is largely determined by the interactions with 
the near-inertial waves. 
\item The accumulation of energy of the near-inertial waves
and the spectrum in the inertial subrange 
could  qualitatively be described by ``named'' nonlocal interactions. 
\end{itemize}

The stratified rotating turbulence governed by the Navier--Stokes equation
and its relation with the
ocean will of course be further researched and debated.
Here we address the reduced wave-wave numerical model of stratified rotating turbulence
with the
hope that it could shed light on processes in oceans and its
spectral energy density in particular.  
It appears that this reduced model reproduce key features of spectral energy density 
of oceanic internal waves. This supports observationally-based intuition that the spectral energy 
density of internal waves is described predominantly by the wave-wave interactions.
The future will certainly
bring many further exciting developments, and the synthesis of theoretical,
observational and numerical results yet to be obtained.

\begin{acknowledgments}
This research is supported by NSF CMG grant 0417724.
Y.~L. was also supported by NSF CAREER DMS 0134955.
We are grateful to YITP in Kyoto University for allowing us to use SX8,
where numerical simulations were performed. We thank Kurt Polzin and Esteban Tabak for multiple and fruitful discussions.
\end{acknowledgments}

\end{document}